\documentclass[14pt,twocolumn]{revtex4}
\usepackage{amsmath,graphicx}

\def\U#1{{%
\def\O{\mbox{O}}
\def\u{\mbox{u}}
\mathcode`\u=\mu
\mathcode`\O=\Omega
{\mathrm {#1}}}}
\def\sub#1{_{\scriptsize\mbox{#1}}}

\usepackage{hyperref}
\begin{document}

\title{Demonstration of negative group delays
in a simple electronic circuit}
\author{T. Nakanishi, K. Sugiyama, and M. Kitano}
\affiliation{Department of Electronic Science and Engineering,
Kyoto University\\
Kyoto 606-8501, Japan}
\date{\today}

\begin{abstract}
\vspace{0.5cm}
We present a simple electronic circuit which produces
negative group delays for base-band pulses.
When a band-limited pulse is applied as the input,
a forwarded pulse appears at the output.
The negative group delays in lumped systems share the same mechanism
with the superluminal light propagation,
which is recently demonstrated in an absorption-free,
anomalous dispersive medium [Wang {\it et al.}, Nature {\bf 406}, 277 (2000)].
In this circuit, the advance time 
more than twenty percent of the
pulse width can easily be achieved.
The time constants, which can be in the order of seconds,
is slow enough to be observed with the naked eye
by looking at the lamps driven by the pulses.
\end{abstract}

\maketitle

\section{Introduction}\label{sec:intro}

Brillouin and Sommerfeld investigated the propagation of light pulse
in a dispersive medium described by the Lorentz model.
They pointed out that in the region of anomalous dispersion
the group velocity $v\sub{g}$ could be larger than $c$,
the light velocity in a vacuum,
or even be negative \cite{Brillouin}.
The anomalous dispersion occurs
near the center of absorption lines.
In the case of superluminal group velocities ($v\sub{g}>c$),
the transit time of the light envelope 
through the medium is shorter than that for a vacuum with the 
same length.
In the case of negative group velocities ($v\sub{g}<0$),
the envelope leaves before it enters into the medium.

Chu and Wong demonstrated experimentally
that the light in a Ga:N crystal can be
propagated at such extraordinary group velocities \cite{Chu}.
But in these kind of experiments 
it is inevitable that the shape of the light pulse is largely
distorted due to the absorption.
The standard definition of the group velocity, $v\sub{g}=d \omega / d k$,
which describes the velocity of the light envelope or peak,
tends to lose its physical meaning 
under the strong distortion of the waveform.
A new definition of group velocities,
considering the reshape of the spectrum caused by the absorption,
was proposed recently \cite{Tanaka}.

Wang {\it et al.}\ realized anomalous dispersion without absorption
using the gain-assisted linear anomalous dispersion,
and demonstrated the light propagation at the negative group velocities
\cite{Wang,Dogariu}.
For the light pulse propagation in an absorption-free medium,
the normal definition of the group velocity describes precisely
the propagating speed of the waveform, or the envelope.

The pulse propagation with superluminal or negative group velocities
is very counterintuitive
and is liable to cause many misunderstandings.
However, it is the direct results of the interference of the wave
and is consistent with the relativistic causality.

The anomalous dispersion with no absorption
induces phase shifts depending on the frequencies,
thereby enhances the front part of the pulse by constructive 
interference and cancels the rear part by the destructive 
interference.
The pulse shape is conserved if the phase shift is 
a linear function of the frequency.

\begin{figure}[b]
\begin{center}
 \includegraphics[scale=0.68]{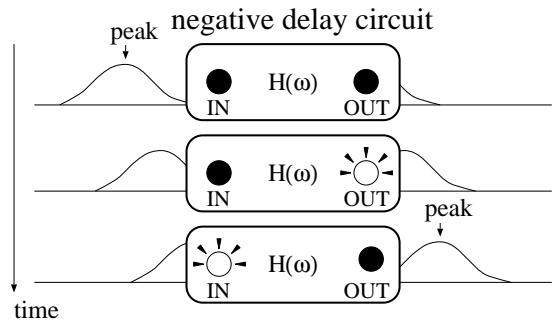}
 \caption{\label{Fig:intro}
An electronic black box for negative delays.
The input and the output are monitored by LEDs.
When a pulse is fed to the input,
the LED at the output lights up before the LED at the input.
}
\end{center}
\end{figure}

The interference does not occur only in the light propagation,
but also in other wave or oscillation dynamics.
Signals in electrical circuits also interfere.
In lumped systems, we cannot define the group velocities,
because there exists no finite length scale.
Instead, we can define the group delay.
The group delay is the time difference between
the input and output signal envelopes.
If it is negative, the output pulse precedes the input as 
shown in Fig.~\ref{Fig:intro}.
The negative delays are closely connected to the superluminal or
negative group velocities in spatially extended systems.

In this paper, we propose a simple electronic circuit which
generates negative group delays.
While the dispersion relation $k(\omega)$ of 
dispersive media determines
the phase shifts in the light propagation,
the transfer function $H(\omega)$ of the circuit
determines the phase shifts of the output
in reference to the input.
Mitchell and Chiao \cite{Mitchell} demonstrated negative delays
in a bandpass filter circuit to which
the carrier signal modulated with a pulse is applied.
They showed that 
if the carrier frequency is located outside of the pass band,
the envelope of the output pulse precedes
the input.
In practical implementation, a large inductor is required and
it is replaced by a simulated inductor with operational amplifiers.
The use of the carrier signal also increases the complication of the
circuit.

The circuit that we present here generates negative delays for 
baseband signals or signals with zero carrier frequency.
It is easy to produce negative delays as large as a few seconds,
therefore, we can observe it with the naked eye
by watching two LEDs (light emitting diodes) each of which is
driven by the input and output pulses.
It is helpful and illuminating to translate fundamental physical concepts
into a simple circuit which replicates 
the essence of the phenomena \cite{Rosner,Frank}.

In the next section
we explain the relation between the superluminal or negative group velocity
in the light propagation and 
the negative group delay in lumped circuits.
Then, in Sec.~\ref{sec:circuit},
we present a circuit generating negative group delays and 
the pulse generator used for the input.
In Sec.~\ref{sec:multi},
we propose a method for larger negative delays
by cascading the circuits.
It is shown that the negative delay time can be increased as $\sqrt{n}$ 
without introducing additional distortion of the pulse shape, where 
$n$ is the number of the stages.
But for large $n$, the circuit becomes susceptible to the noise.

\section{Transfer function for negative group delay}
\label{sec:theory}

In this section we deal with a transfer function $H(\omega)$
in order to study negative group delays.
The discussion can be applied to both electric circuits and light
propagation in a dispersive medium.
A band-limited input signal $E\sub{in}(t)$ can be
expressed as a product of
the carrier $\exp (i\omega_0t)$ and the envelope $\mathcal{E}\sub{in}(t)$;
$E\sub{in}(t) = \mathcal{E}\sub{in}(t) \, e^{i \omega_0 t} + {\rm
c. c. }$,
where ${\rm c.c.}$ represents the complex conjugate term.
Our circuit deals with the baseband signal ($\omega_0 = 0$),
but we include the carrier for the purpose of comparison with
other cases.
The signal is expanded with the Fourier component
$\tilde{\mathcal{E}}\sub{in}(u)$ of 
the envelope ${\mathcal{E}}\sub{in}(t)$ as
\begin{align}
 E\sub{in}(t) = \int^{\Omega/2}_{-\Omega/2} du \,
 \tilde{\mathcal{E}} \sub{in}(u) \, e^{i (\omega_0+u) t} 
 + {\rm c. c. } , \label{Eq:input}
\end{align}
where $\Omega$ is the bandwidth and $u$ is the offset frequency from 
$\omega_0$.

The transfer function $H(\omega) \equiv A(\omega) \, e^{i \phi(\omega)}$
is defined for each frequency $\omega=\omega_0+u$ and
the output $E\sub{out}(t)$ can be written as 
\begin{align}
 E\sub{out}(t) &= 
\int^{\Omega/2}_{-\Omega/2} du \,
 \tilde{\mathcal{E}} \sub{in}(u) H(\omega) \, e^{i (\omega_0+u) t} 
 + {\rm c. c. } \nonumber \\
 &=
\int^{\Omega/2}_{-\Omega/2} du \,
 \tilde{\mathcal{E}} \sub{in}(u) A(\omega) \, e^{i (\omega_0+u) t} 
 e^{i \phi(\omega)}
 + {\rm c. c. } 
\label{Eq:output} 
\end{align}

We assume that
within the bandwidth ($|u|<\Omega/2$),
the amplitude $A(\omega)$ is nearly unity and the phase $\phi(\omega)$ 
can be approximated by a linear function, i.e.,
\begin{align}
A(\omega)\sim 1,\quad
\phi(\omega) \sim \phi(\omega_0) -  u \, t\sub{d} . \label{Eq:phi}
\end{align}
The group delay $t\sub{d}$ is defined as
\begin{align}
 t\sub{d} = - \frac{d \phi}{d \omega} \Big|_{\omega_0} . \label{Eq:delay}
\end{align}
Then the envelope of the output 
is obtained from Eqs.~(\ref{Eq:output}) and (\ref{Eq:phi}) as
\begin{align}
 \mathcal{E}\sub{out}(t) = \mathcal{E}\sub{in}(t-t\sub{d}) 
 \, e^{i \phi(\omega_0)} . \label{Eq:in_out}
\end{align}
This means that, aside from the phase factor,
the envelope of the output is
shifted by the group delay $t\sub{d}$, while maintaining the shape.
For $t\sub{d}>0$,
the input precedes the output (normal delay)
and for $t\sub{d}<0$, the output precedes the input (negative delay).
We will represent a circuit that satisfies the latter condition 
in the next section.

In order to translate the above discussion into the light propagation
through a dispersive medium with length $L$,
we suppose $E\sub{in}$ and $E\sub{out}$ represent 
the field of input and that of output of the medium, respectively.
When the monochromatic light with frequency $\omega$ is propagated 
in the medium,
the phase of the field is shifted as
$\phi(\omega) = - k(\omega) L$,
where $k(\omega)$ is the wavenumber in the medium.
If $k(\omega)$ is linear in the bandwidth, 
with the help of Eq.~(\ref{Eq:delay}),
we have
\begin{align}
 t\sub{d} = L \frac{d k}{d \omega}\Big|_{\omega_0} = \frac{L}{v\sub{g}} ,
 \label{Eq:group_velocity}
\end{align}
where $d \omega/ d k |_{\omega_0}$ is the group velocity $v\sub{g}$.
The envelope of the light is delayed by $L/v\sub{g}$.

If the difference between the propagation time of the envelopes
in the dispersive medium and that in a vacuum with the same length $L$
is negative, i.e.,
\begin{align}
\Delta t = t\sub{d} - L / c=L(v\sub{g}^{-1}-c^{-1})<0,
\end{align}
then the light propagation in the medium is called superluminal.
There are two cases which satisfy this condition;
$v\sub{g}>c$ and $v\sub{g}<0$.
In the former case, the output envelope precedes 
the output for the vacuum case but does not precede the input.
In the latter case, the output precedes the input, or
the whole system provides the negative group delay $t\sub{d}<0$.

\section{Circuits and experiment}
\label{sec:circuit}

\subsection{\label{sub:advance_circuit}
Negative delay circuit for baseband signals}

\begin{figure}[]
\begin{center}
 \includegraphics[scale=1.2]{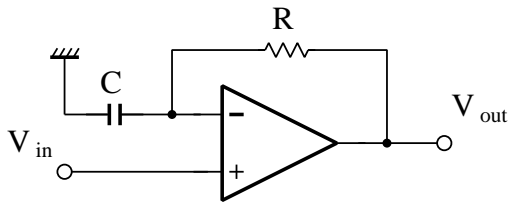}
 \caption{\label{Fig:advance_circuit}
Negative delay circuit for baseband signal can be constructed
with an operational amplifier.
The transfer function is $H(\omega)=1+i\omega T$.
}
\end{center}
\end{figure}

In Fig.~\ref{Fig:advance_circuit},
we show a negative delay circuit for baseband ($\omega_0=0$) signals.
This is basically a non-inverting (imperfect) differentiator.  
Its transfer function is easily obtained as
\begin{align}
 H(\omega) = A(\omega) e^{i \phi(\omega)}
 = 1 + i \omega T, \label{Eq:H_single}
\end{align}
where $T=RC$ .

We note that the corresponding action in the time domain is $1+T(d/d t)$.
The advancement of a pulse can be understood qualitatively.
For the rising edge, which has a positive slope,
the two terms interfere constructively, while for
the falling edge with a negative slope, they interfere destructively.
Thus the pulse is forwarded.

In the low-frequency region ( $|\omega| \ll 1/T$ ),
$H(\omega)$ is approximated as
\begin{align}
 A(\omega)    &= 1 + O(\omega^2 T^2), \label{Eq:amp_single} \\
 \phi(\omega) &= \omega T + O(\omega^3 T^3), \label{Eq:phase_single}
\end{align}
which mean that the amplitude is nearly constant and the phase 
increases linearly with frequency.
Then the group delay becomes negative:
\begin{align}
 t\sub{d} = - \frac{d \phi}{d \omega} \Big|_{\omega=0} = -T <0.
 \label{Eq:advance_time}
\end{align}

\begin{figure}
\begin{center}
 \includegraphics[scale=0.68]{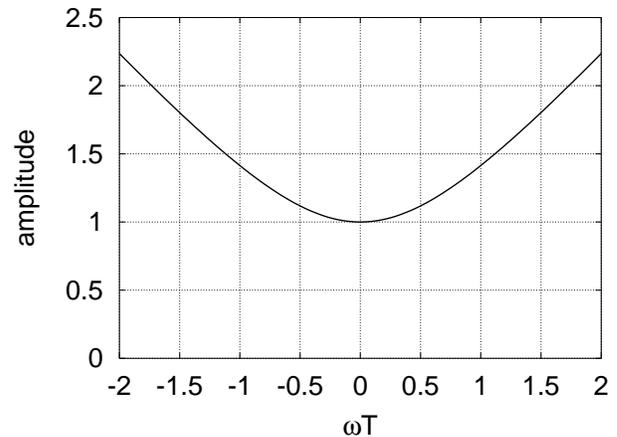}
 \includegraphics[scale=0.68]{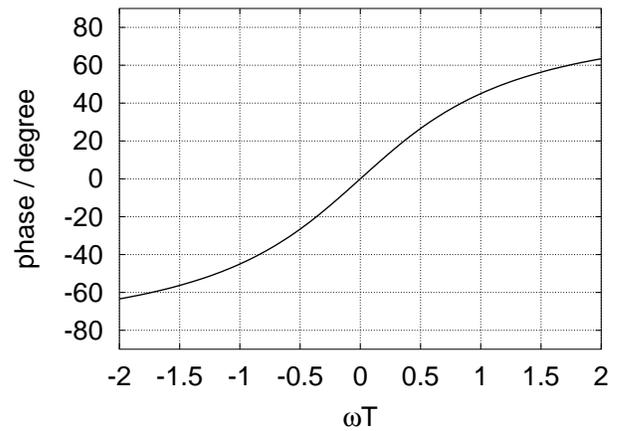}
 \caption{\label{Fig:function}
 The amplitude of the transfer function 
 $A(\omega)=|H(\omega)|$ (upper) and the phase 
 $\phi(\omega)={\rm arg}\, H(\omega)$ (lower).}
\end{center}
\end{figure}

As seen in Fig.~\ref{Fig:function},
the amplitude $A(\omega)$ and the phase $\phi(\omega)$ of 
the transfer function are not linear except in $|\omega| T \ll 1$ 
owing to the higher order terms in Eqs.~(\ref{Eq:amp_single})
and (\ref{Eq:phase_single}).
These terms induce waveform distortion of the output.
In order to keep the distortion as small as possible,
the spectrum of the input signal must be restricted within the frequency 
region $|\omega| \ll 1/T$.
For this purpose low-pass filters are needed.

\subsection{\label{sub:lowpass}Low-pass filter}

\begin{figure}
\begin{center}
 \includegraphics[scale=0.56]{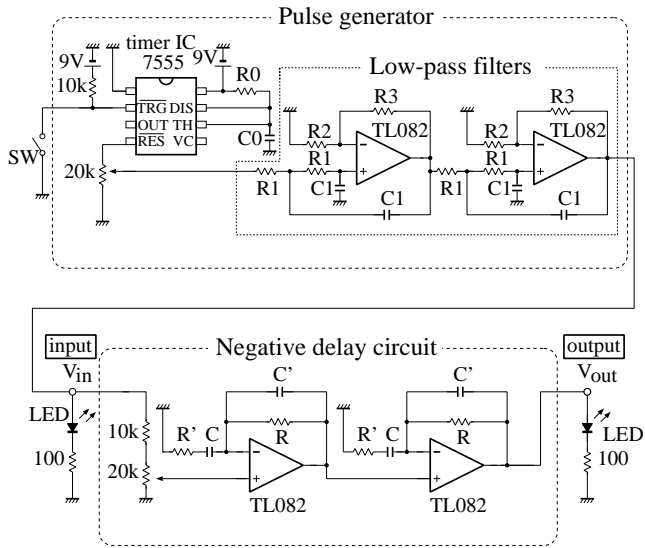}
 \caption{\label{Fig:all_cir}Overall circuits.
Upper section: pulse generator.
Lower section: negative delay circuit.
}
\end{center}
\end{figure}

In order to prepare
a band-limited, base-band pulse we introduce low-pass filters.
The initial source is a rectangular pulse from 
a timer IC (integrated circuit).
The rectangular pulse has high-frequency components,
and the negative delay circuit does not work correctly.
We must eliminate the high-frequency components 
with low-pass filters.
We introduce two 2nd-order low-pass filters shown in Fig.~\ref{Fig:all_cir}.
Each transfer function is
\begin{align}
 H \sub{LP} (\omega) = \frac{\alpha}{1 + i \omega T \sub{LP}
 (3-\alpha) + (i \omega T \sub{LP} )^2}, \label{Eq:Bessel}
\end{align}
where $T\sub{LP} = R_1 C_1$ and $\alpha=(1+R_3/R_2)$. 
By changing $\alpha$,
we can tailor the characteristic of the filter.
In our experiment we choose the value $\alpha=1.268$, 
which corresponds to a Bessel filter \cite{Tietze}.
The Bessel filter is designed so that 
the overshoot for rectangular waves is small.
The cut-off frequency is defined by $\omega\sub{c}=0.7861/T\sub{LP}$ .

\begin{table}
\caption{\label{table:parameters}Circuit parameters.}
\begin{tabular}{ll|ll}
 \hline \hline
   $R_0$        & $6.8 \,\U{M\Omega}$ &
   $R$          & $1 \,\U{M\Omega}$ \\
   $C_0$        & $0.22 \,\U{\mu F}$  &
   $C$          & $0.22 \,\U{\mu F}$  \\
   $T\sub{rec}$ & $1.5 \,\U{s}$     &
   $R^\prime$   & $10 \,\U{k\Omega}$ \\
   $R_1$        & $2.2 \,\U{M\Omega}$ &
   $C^\prime$   & $22 \,\U{n F}$  \\
   $C_1$        & $0.22 \,\U{\mu F}$  &
   $T$          & $0.22 \,\U{s}$    \\
   $R_2$        & $10 \,\U{k\Omega}$  &&\\
   $R_3$        & $2.2 \,\U{k\Omega}$ &&\\
   $\omega\sub{c}$ & $1.6 \,\U{Hz}$   &&\\
 \hline \hline
\end{tabular}
\end{table}

\subsection{\label{sub:exp}Experiment}

We show the overall circuit diagram for
the negative delay experiment in Fig.~\ref{Fig:all_cir} and
the parameters in Table \ref{table:parameters}.

The pulse generator in the upper section of Fig.~\ref{Fig:all_cir}
is subdivided into the generator of single-shot rectangular wave
and the low-pass filters.
In the first part, when triggered by the switch,
the timer IC generates a single pulse, 
whose width is determined by the time constant 
$T\sub{rec}=R_0 C_0=1.5 \,\U{s}$.
The rectangular pulse is shaped by the two-stage low-pass filters.
The total order of low-pass filter is $m=4$.
We set the cut-off frequency of the low-pass filter as $\omega\sub{c}=0.35/T$,
so that $A(\omega)$ and $\phi(\omega)$ can be considered to be constant
and linear, respectively,
below the cut-off frequency
(see Fig.~\ref{Fig:function}).
Finally, the band-limited single pulse is sent out
for the input of the negative delay circuit in the lower section of 
Fig.~\ref{Fig:all_cir}.

Two delay circuits shown in Fig.~\ref{Fig:advance_circuit} are cascaded
for larger advance time.
The input and output terminals are monitored by LEDs.
Their turn-on voltage is about $1.1\U{V}$.
The variable resistor at the input is adjusted so that
the input and the output have the same height.

\begin{figure}[]
\begin{center}
 \includegraphics[scale=0.68]{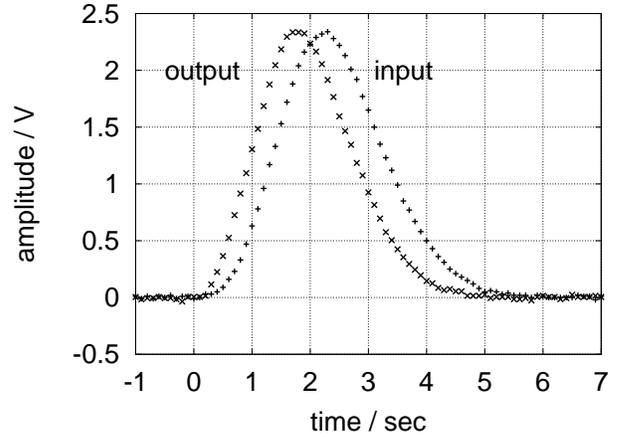}
\caption{\label{Fig:result}Experimental results.
The oscilloscope traces show the input and output pulses.
The output precedes the input owing to the negative delay.}
\end{center}
\end{figure}

The experimental result is shown in Fig.~\ref{Fig:result}.
The input and output waveforms are recorded with a oscilloscope.
The origin of the time ($t=0$) is the moment
when the switch in Fig.~\ref{Fig:all_cir} is turned on.
We see that the output precedes the input considerably
(more than 20\% of the pulse width).
The slight distortion of the output waveform is caused by 
the non-ideal frequency dependence
of $A(\omega)$ and $\phi(\omega)$, as
mentioned in Sec.~\ref{sub:advance_circuit}.

The expected negative delay derived from Eq.~(\ref{Eq:advance_time})
is $2T=0.44\,\U{s}$;
we have connected two circuits
(each time constant $T=0.22\,\U{s}$)
in series for larger effect.
This quantitatively agrees with the experimental result,
where the time difference between the output and the input
peaks is about $0.5\,\U{s}$.
The time scale is chosen so that
one can directly observe the negative delay
with two LEDs connected at the input and the output terminals.
We could also use two voltmeters (or circuit testers) to
monitor the waveforms.
We could dispense with an oscilloscope.

The actual negative delay circuit 
in Fig.~\ref{Fig:all_cir} differs from that shown
in Fig.~\ref{Fig:advance_circuit}.
The resistor $R^\prime$ and the capacitor $C^\prime$
are supplemented for the suppression of 
high-frequency noises.
As shown in Fig.~\ref{Fig:function},
the gain at $|\omega| T>1$ is large.
Although the high-frequency components of the input signal are suppressed
by the low-pass filters,
the internal and external noises 
with high frequency are unavoidable.
With $R'$ and $C'$, the high-frequency gain
is limited.
The parameters are chosen as $R'C , C'R \ll T$ so that the phase
$\phi(\omega)$ for $|\omega|T<1$ is not affected.

\section{Realization of larger negative delay}
\label{sec:multi}

\begin{figure}[]
\begin{center}
 \includegraphics[scale=0.68]{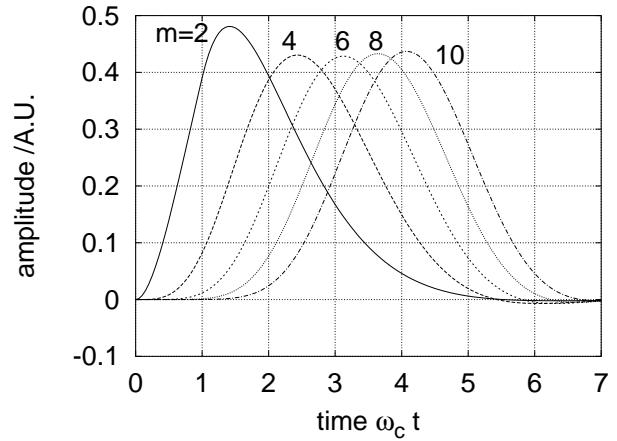}
 \caption{\label{Fig:pulse}Responses of the Bessel filter to a rectangular
 wave. The orders of filters are $m=2,\, 4,\, 6,\, 8,\; \mbox{and} \; 10$.}
\end{center}
\end{figure}

The obtained negative delay in the above experiment is about 20\%
of the pulse width.
This is larger than the values obtained in other 
superluminal-velocity
or negative-delay experiments.
We consider the way to make even larger negative delays.
We assume noise-free environment for a simplicity.

Cascading the negative delay circuits,
the time advance can be increased.
One might simply expect that,
by increasing the number of stages $n$,
the total time advance can be
increased linearly with $n$.
But, unfortunately, this is not the case.
It is obvious from Eqs.~(\ref{Eq:amp_single}) 
and (\ref{Eq:phase_single}) 
that the distortion of the waveform of the output is also increased.
When $n$ circuits are connected in cascade,
the total transfer function can be written as $H^n(\omega)$.
Correspondingly,
the amplitude and the phase are given as
\begin{align}
 A^n(\omega) &\sim 1 + \frac{n(\omega T )^2}{2}, \label{Eq:amp_multi} \\
 n \phi(\omega) &\sim n \omega T. \label{Eq:phase_multi}
\end{align}
In order to keep the wave distortion below a certain level,
we have to limit the excess gain $A^n(\omega)-1$ within 
the bandwidth by some value $\gamma$;
\begin{align}
\frac{n(\omega T )^2}{2} \leq 
\frac{n(\omega\sub{c} T )^2}{2} = \gamma.
\label{Eq:gain_limit}
\end{align}
Then the advance time per circuit should satisfy
\begin{align}
T = \sqrt{\frac{2\gamma}{n}}\omega\sub{c}^{-1}
=\sqrt{\frac{2\gamma}{n}}T\sub{w},
\label{Eq:gain_limit}
\end{align}
where $T\sub{w}$ is the pulse width.
It is determined by the cut-off frequency of the low-pass filter.

If we want to increase the time advance in conserving the pulse width
and the distortion of the signal,
we must reduce the time advance $T$ per circuit by 
the factor $1/\sqrt{n}$.
Therefore, the total time advance $T_{\rm total}$ scales as
\begin{align}
 T_{\rm total} = n T = \sqrt{2 n \gamma}\, T\sub{w}, \label{Eq:multi_advance}
\end{align}
which is a slowly increasing function of $n$.

In addition, there is another factor to be considered.
A causal transfer function cannot generate negative delays
unconditionally.
It is impossible to advance the signal beyond 
the time when the switch is turned on
in the rectangular pulse generator in Fig.~\ref{Fig:all_cir}.
The reason of the advancement
is that the slow rising part of the pulse,
which has been suppressed by the low-pass filter,
is reemphasized by the negative delay circuit.
The slowness of rising part of the pulse is determined 
by the order $m$ of 
the low-pass filter.
Figure \ref{Fig:pulse} represents responses of various order Bessel
filters with the same cut-off $\omega\sub{c}$  
to a rectangular wave
with a unit height and a pulse width $\omega\sub{c}^{-1}$.
The pulse shapes, especially the widths of the pulses, 
are similar to each other owing to the same cut-off frequency,
but the higher the order of filter the more the rising part is delayed.
Hence we need a high order filter to attain large time advance.
In other words, the pulse must be delayed appropriately in advance
in order to get a large negative delay.

Moreover the short-time behavior of the total circuit including 
the low-pass filters and the negative delay circuits is determined by
the composite transfer function at high frequency 
($\omega \rightarrow \infty$).
The order of the low-pass filter $m$ should not be smaller than 
the number of the stages $n$ of the negative delay circuits.
Otherwise the total transfer function would diverge at 
$\omega \rightarrow \infty$ and the derivative of the rectangular
pulse would appear at the output.

\begin{figure}[]
\begin{center}
 \includegraphics[scale=0.68]{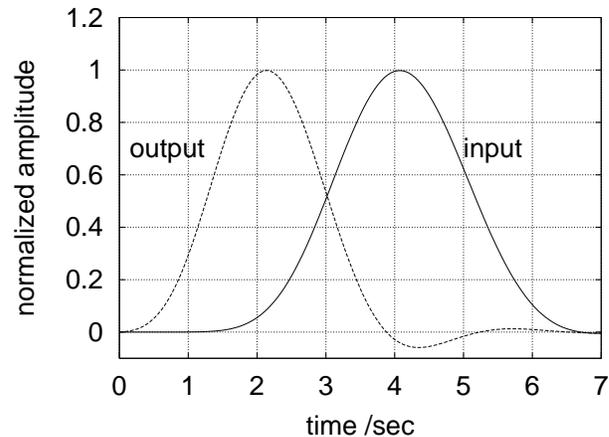}
 \caption{\label{Fig:multi_result}A simulation result for multi-stage
negative delay circuit ($n=10$).}
\end{center}
\end{figure}

Figure \ref{Fig:multi_result} represents a result of simulation
for the response of ten negative delay circuits 
with the input of a tenth order Bessel filter.
The used parameters are $T\sub{w}=1\,\U{s}$ and $\gamma=0.2$.
The total time advance is estimated $2\,\U{s}$
from Eq.~(\ref{Eq:multi_advance}).
This estimation is consistent with the result of the simulation.
The advance time is comparable to the pulse width.
It is so large that
the input starts to rise when the output begins to fall.

\section{Discussion and conclusion}
\label{sec:conclusion}

Let us consider a positive delay circuit.
A circuit called all-pass filter has the transfer function \cite{Tietze}:
\[
H\sub{P}(\omega)=\frac{1-i\omega T}{1+i\omega T}.
\]
The all-pass filter can be build with an operational amplifier,
three resistors, and a capacitor.
It is very convenient for generating positive delays 
for base-band signals because the
amplitude and phase of the transfer functions are given as
$A(\omega)=1$, and $\phi(\omega)=-2\tan^{-1}\omega T\sim -2\omega T$,
respectively.
The flat amplitude response allows us to cascade the
$n$ circuits without introducing pulse distortion and
results in the delay $t\sub{d}=2nT$.
Unfortunately the negative version ($T\rightarrow -T$) of all-pass filter
\[
H\sub{N}(\omega)=\frac{1+i\omega T}{1-i\omega T}
=\frac{H(\omega)}{1-i\omega T},
\]
will not work because it is not causal.
The pole of $H\sub{N}(-is)$, $s=1/T$, is located in the right half plain.
If one makes this circuit, it will be unstable owing to the
time response function $\exp(t/T)$.
This example tells us about the asymmetry between the negative delay and
the positive delay.
The former is much more difficult to achieve than the latter.

The group velocity has no direct connection with the relativistic
causality, therefore, it can exceed the speed of light $c$ in a vacuum.
But the front velocity $v\sub{f}$ (or the wavefront velocity)
is constrained by the causality and is equal to $c$, namely,
$|v\sub{f}|=L/|t\sub{f}| = c$.
In lumped systems ($L=0$), the wavefront delay $t\sub{f}$
must vanish.
Actually all of the pulses in our system have their wavefronts at
$t=0$, the moment when the original rectangular pulse rises or
the switch is turned on.

To conclude, we demonstrated the negative group delay
in a simple electronic circuit.
A considerably large negative delay ($0.44\,\U{s}$) could easily be
achieved.
It is slow enough to be observed with the
LEDs or voltmeters with the naked eye.
The negative delay amounts to 20\% of the pulse width.
In the light experiment \cite{Wang},
$t\sub{d}-L/c$ is $62\,\U{ns}$ and is only a few percents of the pulse
width, $4\U{\mu s}$.

Including the pulse generator for the input of the negative delay circuit,
the apparatus consists of common parts available in any laboratories.
It is so simple that a beginner can build it in an hour.
The setup can be operated stand-alone and
no expensive instruments such as oscilloscopes
and function generators are required.
It is useful for understanding the physics of 
superluminal propagation as well as the negative group delays.

\acknowledgements

This research was supported by the Ministry of
Education, Culture, Sports, Science and Technology
in Japan under a Grant-in-Aid for Scientific Research
No.~11216203 and No.~11650043.


\begin{thebibliography}{}

 \bibitem{Brillouin}
 L. Brillouin,
 Wave Propagation and Group Velocity (Academic Press, New York, 1960)
 pp. 113 -- 137.
 
 \bibitem{Chu}
 S. Chu and S. Wong,
 ``Linear Pulse Propagation in an Absorbing Medium,''
 Phys. Rev. Lett. {\bf 48}, 738 -- 741 (1982).

 \bibitem{Tanaka}
 M. Tanaka, M. Fujiwara, and H. Ikegami,
 ``Propagation of a Gaussian wave packet in an absorbing medium''
 Phys. Rev. A {\bf 34}, 4851 -- 4858 (1986).

 \bibitem{Wang}
 L. J. Wang, A. Kuzmich, and A. Dogariu,
 ``Gain-assisted superluminal light propagation,''
 Nature {\bf 406}, 277 -- 279 (2000).

 \bibitem{Dogariu}
 A. Dogariu, A. Kuzmich, and L. J. Wang,
 ``Transparent anomalous dispersion and superluminal light-pulse 
 propagation at a negative group velocity,''
 Phys. Rev. A {\bf 63}, 053806-1 -- 053806-11 (2001).

 \bibitem{Mitchell}
 M. W. Mitchell and R. Y. Chiao, 
 ``Causality and Negative Group Delays in a Simple Bandpass Amplifier,''
 Am. J. Phys. {\bf 66} (1), 14--19 (1998).

\bibitem{Rosner}
 J. L. Rosner,
 ``Tabletop time-reversal violation,''
 Am. J. Phys. {\bf 64} (8), 982--985 (1996).

 \bibitem{Frank}
 W. Frank and P. Brentano, 
 ``Classical analogy to quantum mechanical level repulsion,''
 Am. J. Phys. {\bf 62} (8), 706--709 (1994).

 \bibitem{Tietze}
 U.Tietze and Ch. Schenk,
 Electronic Circuit (Springer-Verlag, Berlin, 1991)
 pp. 350--408.

\end{thebibliography}
\end{document}